\documentclass[a4paper,12pt]{article}
\usepackage{amsmath}
\usepackage{amssymb}
\usepackage{amsfonts}
\usepackage{bbm} 
\usepackage{fleqn} 
\usepackage[footnotesize]{caption}
\usepackage{graphicx} 
\usepackage{braket} 
\usepackage{mathrsfs}
\usepackage[center,footnotesize,hang]{subfigure}

\newcommand{\CenterEps}[2][1]{\ensuremath{\vcenter{\hbox{\includegraphics[scale=#1]{#2.eps}}}}}

\def\I{\mathrm{i}}

\def\eps{\epsilon}

\def\<{\left\langle}
\def\>{\right\rangle}

\begin{document}

\bibliographystyle{OurBibTeX}

\begin{titlepage}

 \vspace*{-15mm}
\begin{flushright}
SHEP/0524\\
hep-ph/0507333
\end{flushright}
\vspace*{5mm}

\begin{center}
{
\sffamily
\LARGE
Leptogenesis in Unified Theories 
with Type II See-Saw
}
\\[12mm]
S.~Antusch\footnote{E-mail: \texttt{santusch@hep.phys.soton.ac.uk}},
S.~F.~King\footnote{E-mail: \texttt{sfk@hep.phys.soton.ac.uk}}
\\[1mm]
{\small\it
School of Physics and Astronomy,
University of Southampton,\\
Southampton, SO17 1BJ, U.K.
}
\end{center}
\vspace*{1.00cm}

\begin{abstract}

\noindent 
In some classes of flavour models based on unified theories with a type I
see-saw mechanism, 
the prediction for the mass of the lightest right-handed neutrino 
is in conflict with the lower bound from the requirement of successful 
thermal leptogenesis. 
We investigate how lifting the absolute neutrino mass scale by 
adding a type II see-saw contribution proportional to the unit matrix  
can solve this problem.
Generically, lifting the neutrino mass scale increases the prediction
for the mass of the lightest right-handed neutrino while the decay asymmetry 
is enhanced and washout effects are reduced, relaxing the lower bound 
on the mass of the lightest right-handed neutrino from thermal leptogenesis. 
For instance in classes of unified theories where the lightest right-handed 
neutrino dominates the type I see-saw contribution, we find that 
thermal leptogenesis becomes possible if the neutrino mass scale is larger than  
about 0.15 eV, making this scenario testable by neutrinoless double 
beta decay experiments in the near future.

\end{abstract}

\end{titlepage}
\newpage
\setcounter{footnote}{0}

\section{Introduction}
The mechanism of Leptogenesis \cite{Fukugita:1986hr} is one of the most attractive
possibilities for explaining the observed baryon asymmetry of the universe, 
$n_\mathrm{B} /n_\gamma 
= (6.5^{+0.4}_{-0.8}) \cdot 10^{-10}$ \cite{Spergel:2003cb}. 
The asymmetry is generated via the out-of-equilibrium 
decay of the same heavy right-handed neutrinos which are responsible for 
generating naturally small neutrino masses in the type I see-saw scenario 
\cite{seesaw}. 
The thermal version of leptogenesis in the so-called strong washout regime is  
thereby virtually independent of initial conditions, since the effect of 
any pre-existing baryon asymmetry or right-handed neutrino abundance is 
washed out by processes in the thermal bath involving the lightest 
right-handed neutrino. 
In the type I see-saw mechanism, thermal leptogenesis
(assuming hierarchical right-handed neutrino masses) puts strong constraints on the parameters 
of the see-saw mechanism.
To start with, the decay asymmetries are bounded from above and best asymmetry 
is achieved for hierarchical
neutrino masses \cite{Hamaguchi01,Davidson:2002qv}. This leads to a lower bound on the masses of the right-handed
neutrinos \cite{Davidson:2002qv}, which amounts about $10^9$ GeV (see e.g.\
\cite{Giudice:2003jh,Buchmuller:2004nz} for recent calculations) for hierarchical neutrino masses and
increases strongly as the neutrino mass scale increases. Together with the observation
that in the type I see-saw scenario, the washout parameter $\widetilde m_1$
\cite{Buchmuller:2000as} increases with increasing neutrino mass scale
leading to strongly enhanced washout which makes leptogenesis less efficient, a bound
on the absolute neutrino mass scale of about $0.1$ eV can be derived 
\cite{Buchmuller:2003gz}.

In models with a left-right symmetric particle content like minimal left-right 
symmetric models, Pati-Salam models or Grand Unified Theories (GUTs) 
based on SO(10), the type I see-saw mechanism is typically   
generalized to a type II see-saw \cite{seesaw2}, 
where an additional direct 
mass term $m_{\mathrm{LL}}^{\mathrm{II}}$ for the light neutrinos is present. 
The effective mass matrix of the light neutrinos is then given by 
\begin{eqnarray}\label{eq:SeeSawFormula}
m^{\nu}_{\mathrm{LL}} = 
m_{\mathrm{LL}}^{\mathrm{I}}+ m_{\mathrm{LL}}^{\mathrm{II}}\;,\;\;
\mbox{where}\;\;
m_{\mathrm{LL}}^{\mathrm{I}} = - v^2_\mathrm{u}\, 
Y_\nu \, M_\mathrm{RR}^{-1}\, Y_\nu^T 
\end{eqnarray} 
is the type I see-saw mass matrix with $Y_\nu$ being the neutrino Yukawa matrix in
left-right convention, 
$M_\mathrm{RR}$ the mass matrix of the right-handed neutrinos and
$v_\mathrm{u} = \< h^0_\mathrm{u} \>$
is the vacuum expectation value (vev) which leads to masses for the 
up-type quarks.  
From a rather model independent viewpoint, the type II mass term can 
be considered as an additional contribution to the lowest dimensional 
effective neutrino mass operator. In most explicit models, the type II contribution
stems from see-saw suppressed induced vevs of 
SU(2)$_{\mathrm{L}}$-triplet Higgs fields. 

Leptogenesis in type II see-saw scenarios
\cite{Hambye:2003ka,Antusch:2004xy,TypeIILGModels} via the decay of the lightest right-handed
neutrino provides a natural generalization of type I leptogenesis. In the limit that
the mass of the lightest right-handed neutrino is much lighter than the other
particles participating in the see-saw mechanism, the decay asymmetry depends just 
on the low energy neutrino mass matrix 
$m^{\nu}_{\mathrm{LL}} = m_{\mathrm{LL}}^{\mathrm{I}} +
 m_{\mathrm{LL}}^{\mathrm{II}}$ and on the Yukawa
couplings to the lightest right-handed neutrino and its mass 
\cite{Antusch:2004xy}. It has been shown that type II
leptogenesis puts constraints on the see-saw parameters as well, which however differ
substantially from the constraints in the type I case. For instance, 
the bound on the 
decay asymmetry increases with increasing neutrino mass scale 
\cite{Antusch:2004xy}, in contrast to the
type I case where it decreases. As a consequence, the lower bound on the mass of the
lightest right-handed neutrino from leptogenesis decreases for increasing neutrino 
mass scale \cite{Antusch:2004xy}. 
Finally, since the type II contribution typically does not  
effect washout, there is no bound on the absolute 
neutrino mass scale 
in type II leptogenesis \cite{Hambye:2003ka}.

One potential problem for thermal leptogenesis emerges in some classes
of unified theories, where the neutrino Yukawa couplings are linked to
the Yukawa couplings in the up-quark sector which implies
small Dirac mixing from the Yukawa matrices
\cite{MRproblem,Akhmedov:2003dg}.  
In such models the masses of the right-handed
neutrinos calculated within the type I see-saw mechanism are required
to be strongly hierarchical, and the lightest right-handed
neutrino turns out to be so light that it can be below  
the leptogenesis bound of $10^9$ GeV. 
Within the type I see-saw mechanism, proposed solutions to
this potential problem include nearly degenerate right-handed
neutrinos leading to resonant leptogenesis
\cite{Flanz:1996fb,Pilaftsis:2003gt,Albright:2003xb}, non-thermal
leptogenesis via the decay of the inflaton
\cite{nonthermalLG,SneutrinoInflation}, or of course applying a
completely different baryogenesis mechanism. The fact that the type
II see-saw mechanism has the potential for solving the right-handed
neutrino problems in unified theories was also mentioned in
\cite{Akhmedov:2003dg}, but not discussed in any detail.

In this paper we consider realistic classes of unified theories,
where the lightest right-handed neutrino dominates the type I see-saw
mechanism \cite{SequentialD}.  We show that in this scenario, 
the prediction for the mass of the lightest right-handed neutrino is
generically in conflict with the lower bound from the requirement of
successful thermal leptogenesis. Although our predictions for the
masses of the right-handed neutrinos are somewhat larger than the
estimated range given in \cite{Akhmedov:2003dg}, we show that in such 
models leptogenesis is strongly washed out, leading to a more 
stringent lower limit on the right-handed neutrino mass
of about $10^{11}$ GeV which is in conflict with the 
allowed range of right-handed neutrino masses from the class
of unified model considered. 

The main purpose of this paper is to investigate how
lifting the absolute neutrino mass scale by adding a type II see-saw
contribution proportional to the unit matrix can lead to a resolution 
of the conflict between the leptogenesis lower bound on the 
lightest right-handed neutrino mass, and the allowed range
of lightest right-handed neutrino masses in classes of unified
theories where the neutrino Yukawa couplings are related to the
up-quark ones. We have previously shown that 
such a ``type II upgrade'' \cite{Antusch:2004xd} provides a
natural way for transforming a type I see-saw model for hierarchical
neutrino masses into a type II see-saw model for quasi-degenerate
neutrinos \cite{Caldwell:1993kn}. 
Increasing the neutrino mass scale
using the type II see-saw mechanism implies that the mass splittings
between the physical neutrino masses are reduced, and since 
in this approach these
splittings are controlled by the type I see-saw mechanism, this 
has the effect of increasing the masses of the right-handed neutrinos
required to give a successful description of neutrino masses.
Increasing the type II contributions also implies that 
the decay asymmetries become larger and washout effects are
reduced, which reduces the lower bound on the mass of the lightest right-handed
neutrino from thermal leptogenesis. The combination of these two
effects implies that, as the type II neutrino mass scale
increases, the increasing lightest right-handed neutrino mass 
prediction converges with the decreasing leptogenesis lower limit, 
thereby resolving the
conflict between unified theories and thermal leptogenesis.
Quantitatively we find that the conflict is resolved for 
a neutrino mass scale larger than about 0.15 eV. Our scheme therefore
predicts a signal in neutrinoless double beta decay experiments 
(and possibly also in direct searches for neutrino mass) 
\cite{Aalseth:2004hb,Klapdor-Kleingrothaus:2004wj} in the near future.    
An additional nice feature of our proposal is that 
for such neutrino masses, thermal leptogenesis remains in the
   so-called strong washout regime, where the produced baryon
   asymmetry is virtually independent of initial conditions.

It is worth mentioning that 
an analogous problem appears in unified theories where the lightest
right-handed neutrino determines the sub-dominant contributions to the
neutrino mass matrix, and thermal leptogenesis requires a similar lift
of the neutrino mass scale in this case. 
%
On the contrary, the above-mentioned conflict is typically absent, if
the heaviest right-handed neutrino is dominant
\cite{SequentialD,Hirsch:2001dg} and it can be ameliorated if the
dominance conditions are relaxed \cite{Dermisek:2004tx}.

\section{Leptogenesis in Unified Theories with Type I See-Saw}

\subsection{Unified Models with Dominant Lightest RH Neutrino $\boldsymbol{\nu_\mathrm{R1}}$}
In order to discuss predictions for right-handed neutrino masses and
issues of thermal leptogenesis explicitly, it is necessary to make
assumptions.
We will therefore consider first a class of unified models motivated
by left-right symmetric unified theories such as GUTs based on SO(10),
where the lightest right-handed neutrino $\nu_\mathrm{R1}$
dominates the see-saw
mechanism.  In these classes of models we are led to specific forms of
the Yukawa couplings, which we will now briefly review. More details
and explicit examples for models within this class of unified flavour
models can be found in Ref.~\cite{King:2001uz}.  

The known
experimental data about fermion masses and mixings can be successfully
accommodated by Yukawa matrices for up-type quarks $Y_\mathrm{u}$,
down-type quarks $Y_\mathrm{e}$, charged leptons $Y_\mathrm{d}$ and
neutrinos $Y_\nu$ all being of the form
\begin{eqnarray}
Y_f \sim 
\begin{pmatrix}
0        & \eps_f^3 & \eps_f^3 \\
\eps_f^3   & x_f\eps_f^2 +\eps_f^3 & x_f\eps_f^2 +\eps_f^3\\
\eps_f^3   & x_f\eps_f^2 +\eps_f^3 & {\cal O}(1) \\
\end{pmatrix} ,
\label{Yf}
\end{eqnarray} 
where $f = \mathrm{u},\mathrm{d},\mathrm{e},\nu$ and where in the
quark sector $\eps_{\mathrm{u}} \approx 0.05$ and $\eps_{\mathrm{d}}
\approx 0.15$ are different expansion parameters obtained from a fit
to $\frac{m_\mathrm{c}}{m_\mathrm{t}}$ and
$\frac{m_\mathrm{s}}{m_\mathrm{b}}$. In the charged lepton sector, a
Clebsch factor (Georgi-Jarlskog factor \cite{Georgi:1979df}) of -3 in
the (2,2)-entry of $Y_\mathrm{d}$ is typically introduced and the
expansion parameter $\eps_{\mathrm{e}}$ is equal to
$\eps_{\mathrm{d}}$. More specifically, one might connect the factor
$x_f$ appearing in the above texture to the weak hypercharge,
suggesting $x_{\mathrm{d}} = 1, x_{\mathrm{d}} = -2, x_{\mathrm{e}} =
-3$ and $x_\nu = 0$.  The approximate texture zero in the (0,0)-entry
of $Y_f$ furthermore leads to the successful GST relation \cite{GST},
which relates quark masses and the Cabibbo angle.
Note that the Yukawa matrices in Eq.~(\ref{Yf}) are written in a
left-right convention in which the first column gives the couplings
to the first right-handed fermion, and so on.

Although not unique, the texture in Eq.~(\ref{Yf}) has the feature that
all the charged fermion mixing angles are small, which is common to 
many SO(10) type models. Within this class of models, we shall
obtain large neutrino mixing using a mechanism called light sequential
dominance (LSD), in which the lightest right-handed neutrino dominates
the type I see-saw contribution to the atmospheric mass,
and the next-to-lightest right-handed neutrino dominates the
the type I see-saw contribution to the solar neutrino mass
\cite{SequentialD}. Although the choice of LSD is also not unique,
it has the desirable feature that a neutrino mass hierarchy arises
naturally without any tuning, since the large neutrino mixing angles
are given by ratios of Yukawa couplings, and the problem of large
neutrino mixing is therefore decoupled from the neutrino mass
hiearchy which arises naturally from the sequential dominance
of the three right-handed neutrinos.

\subsection{Estimating 
the Lightest RH Neutrino Mass $\boldsymbol{M_\mathrm{R1}}$ from Unification}\label{sec:TypeI}

In the classes of models outlined above, the neutrino Yukawa matrix has the form  
\begin{eqnarray}\label{eq:YnuLSD}
Y_\nu =
\begin{pmatrix}
0         & a\eps^3 & p\eps^3 \\
e\eps^3   & b\eps^3 & q\eps^3 \\
f\eps^3   & c\eps^3 & {\cal O}(1) \\
\end{pmatrix} ,
\end{eqnarray} 
where $a,b,c,e,f,p,q$ are order unity dimensionless couplings, and
$\eps := \eps_\nu = \eps_\mathrm{u}$.  Note that the entries
proportional to $\eps^2$ are absent due to a vanishing Clebsch factor
$x_\nu = 0$. Providing the lightest right-handed neutrino 
(corresponding to the first column in Eq.~(\ref{eq:YnuLSD}))
provides the dominant type I see-saw contribution to the atmospheric 
neutrino mass, then we are
naturally led to large atmospheric neutrino mixing 
$\tan \theta_{23}\approx e/f \sim 1$ 
for $e \sim f$ and a hierarchical 
neutrino mass spectrum,
even with an approximately diagonal mass matrix
\begin{eqnarray}
M_\mathrm{RR} \;=\; \mbox{diag}\,(M_\mathrm{R1} ,M_\mathrm{R2} ,M_\mathrm{R3} )
\end{eqnarray}
for the right-handed neutrinos. This is the single right-handed
neutrino dominance mechanism \cite{SequentialD}.
Furthermore, if the next-to-lightest right-handed neutrino
(corresponding to the second column in Eq.~(\ref{eq:YnuLSD}))
provides the dominant type I see-saw contribution to the solar neutrino 
mass, then we are naturally led to 
large solar mixing $\tan \theta_{12}\approx \sqrt{2}a/(b-c)$ 
for $a\sim b \sim c$, which is 
the sequential neutrino dominance mechanism \cite{SequentialD}.
In the following, we will assume the sequential dominance
conditions \cite{SequentialD}  
 \begin{eqnarray}\label{Eq:SequentialD}
\frac{|e|^2\eps^6,|f|^2\eps^6}{M_\mathrm{R1}} \;\gg\; 
 \frac{|a|^2\eps^6,|b|^2\eps^6,|c|^2\eps^6}{M_\mathrm{R2}}\;\; \gg 
 \;\;\frac{1}{M_\mathrm{R3}}\; ,
 \end{eqnarray}
which immediately leads to a physical 
neutrino mass hierarchy $m_1\ll m_2 \ll m_3$.
It also implies a hierarchy of heavy right-handed neutrino masses
$M_\mathrm{R1} \ll M_\mathrm{R2}\ll M_\mathrm{R3}$.  
The mass of the lightest right-handed neutrino,
 which dominates the type I see-saw mechanism, is then given by
 \cite{SequentialD}
\begin{eqnarray}\label{eq:MR1inTypeILSD}
M_\mathrm{R1} \;=\; \frac{(e \eps^3)^2 v_\mathrm{u}^2}{(s^\nu_{23})^2 m^\mathrm{I}_3} \; ,
\label{eq:MR1_LSD}
\end{eqnarray}   
where the contribution to the masses of the light neutrinos from the 
type I see-saw mechanism are denoted by $m^\mathrm{I}_i$, $i=1,2,3$,  
in order to distinguish them from the type II
see-saw contribution we will introduce in Sec.~\ref{sec:TypeII}. In the type I see-saw case with
a hierarchical neutrino mass spectrum,
$m_3^\mathrm{I}$ is simply given by $\sqrt{|\Delta m^2_{\mathrm{atm}}|}$ with 
$ |\Delta m^2_{\mathrm{atm}}| 
\approx 2.2 \cdot 10^{-3}$ eV$^2$ \cite{Maltoni:2004ei}. 
Nearly maximal mixing $\theta_{23}$ stems from the neutrino sector, i.e.\
$s_{23} \approx s^\nu_{23}$ with only small
corrections from the charged leptons, and we use 
$s^\nu_{23} = 1/\sqrt{2}$ in the following. 
Furthermore, in the classes of models outlined above, the neutrino Yukawa
matrix $Y_\nu$ is related to the up-type quark Yukawa matrix $Y_\mathrm{u}$,
although it is of course not required that both Yukawa matrices have to be
identical. For estimating $M_\mathrm{R1}$, we are interested in the
(2,1)-entry of $Y_\nu$ which is equal to $e\eps^3$.   
 Explicit fits of the quark sector suggest that 
$(Y_\mathrm{u})_{12} = (Y_\mathrm{u})_{21} = 1.5 \eps_\mathrm{u}^3$ 
\cite{Roberts:2001zy} at
$M_\mathrm{GUT}$, and for our estimates we will allow $(Y_\nu)_{21}$ to
vary from about $\tfrac{1}{5} \cdot (Y_\mathrm{u})_{21}$ to 
$5 \cdot (Y_\mathrm{u})_{21}$. Such differences from the quark Yukawa matrix
might stem from different Clebsch factors and/or from uncertainties in the quark
masses which lead to uncertainties in $\eps$.  
With $e\eps^3 \in [\tfrac{1}{5},5] \cdot 1.5 \eps^3$, we obtain
\begin{eqnarray}\label{eq:EsimatedRangeForMR1}
M_\mathrm{R1} \;\sim \; 2 \cdot 10^6 \:\mbox{GeV} \; ... \; 
1 \cdot 10^9\:\mbox{GeV}\;,
\end{eqnarray}
neglecting RG corrections at this stage. 
The range for $M_{\mathrm{R1}}$ can of course be extended/reduced
somewhat by assuming a larger/smaller range for $e\eps^3$.  Note that
although it seems that the range of Eq.~(\ref{eq:EsimatedRangeForMR1})
is marginally consistent with the absolute lower bound on
$M_\mathrm{R1}$ of about $10^9$ GeV \cite{Davidson:2002qv} from
thermal leptogenesis, we will show below that for the lightest
right-handed neutrino dominating the see-saw mechanism (LSD), this
bound is in fact much more stringent and in gross conflict with the
predicted range for $M_\mathrm{R1}$ of
Eq.~(\ref{eq:EsimatedRangeForMR1}).

\subsection{Lower Bound on $M_\mathrm{R1}$ from Thermal Leptogenesis}
The observed baryon asymmetry of the universe is given by 
$n_\mathrm{B} /n_\gamma 
= (6.5^{+0.4}_{-0.8}) \cdot 10^{-10}$ \cite{Spergel:2003cb}. 
This has to be compared to the baryon-to-photon ratio produced by 
leptogenesis which can be calculated from the formula
(using a notation as, e.g., in \cite{Giudice:2003jh}) 
\begin{eqnarray}\label{eq:nBformEpsEta}
  \frac{n_\mathrm{B}}{n_\gamma}
 &\approx& - \,1.04 \cdot 10^{-2}\,\varepsilon_{1}
 \, \eta\; ,
 \end{eqnarray}
where $\varepsilon_{1} $ is the decay asymmetry of the lightest right-handed 
neutrino into lepton doublet and Higgs and where the 
parameter $\eta$ is the so-called efficiency factor, 
which e.g.\ takes dilution of the produced
asymmetry by washout processes into account.

For the type I see-saw mechanism, the decay asymmetry 
$\varepsilon_{1}$ \cite{Covi:1996wh} in the MSSM can be written as 
\begin{eqnarray}\label{eq:eps1typeI}
\varepsilon_{1}&=& \frac{1}{8 \pi} 
 \frac{\sum_{j\not=1}\mbox{Im}\, [(Y^\dagger_\nu Y_\nu)^2_{1j}]}{
 \sum_f \, |(Y_\nu)_{f1}|^2} 
 \, \sqrt{x_j} \,\left[ \frac{2}{1-x_j} - \ln \left( \frac{x_j +1}{x_j}\right)
 \right] \nonumber \\
 &\approx&
 \frac{-3}{8 \pi} 
 \frac{\mbox{Im}\, [(Y^\dagger_\nu Y_\nu)^2_{12}]}{
 \sum_i \, |(Y_\nu)_{i1}|^2} 
 \, \frac{M_\mathrm{R1}}{M_\mathrm{R2}} 
 =
 \frac{3}{8 \pi} 
 \frac{M_{\mathrm{R}1}}{v_\mathrm{u}^2} 
 \frac{\sum_{fg}\mbox{Im}\, [(Y^*_\nu)_{f1} (Y^*_\nu)_{g1} 
 (m_{\mathrm{LL}}^\mathrm{I})_{fg}]}{(Y_\nu^\dagger Y_\nu)_{11}}
\end{eqnarray}
with $x_j := M^2_{\mathrm{R}j} / M^2_{\mathrm{R}1}$ for $j \not= 1$.
In the second line, we have used that $M_\mathrm{R3}$  
effectively decouples from the see-saw
mechanism and from leptogenesis and that  
$M_\mathrm{R1}\ll M_\mathrm{R2}\ll M_\mathrm{R3}$ (cf.\ Eq.~(\ref{Eq:SequentialD})).

The efficiency factor $\eta$ can be computed from a 
set of coupled Boltzmann equations
 (see e.g.\  \cite{Buchmuller:2000as}) and 
 it is subject to e.g.\ thermal
 correction \cite{Giudice:2003jh} and corrections from spectator processes 
 \cite{Buchmuller:2001sr}, 
 $\Delta\mathrm{L} \!=\! 1$ processes involving gauge bosons 
 \cite{Pilaftsis:2003gt,Giudice:2003jh} and from renormalization group
 running \cite{Barbieri:1999ma}.  
 For $M_{\mathrm{R}1}$ much smaller than $10^{14}$ GeV \cite{Giudice:2003jh}, to 
 a good approximation  
 the efficiency factor depends only on the quantity $\widetilde m_1$ 
 \cite{Buchmuller:2000as}, defined by 
\begin{eqnarray}\label{eq:m1tilde}
\widetilde m_1 := \frac{\sum_{f} (Y^\dagger_\nu)_{1f} (Y_\nu)_{f1}\, 
v_\mathrm{u}^2 }{ M_{\mathrm{R1}} } \; . 
\end{eqnarray} 
For $\widetilde m_1$ larger than about $10^{-2}$ eV, $\eta$ is independent of the 
initial population of right-handed (s)neutrinos (see e.g.\ Fig.~8 of 
\cite{Giudice:2003jh}). In this range, larger 
$\widetilde m_1$ means
larger washout and a reduced efficiency factor $\eta$. 
For $\eta$, we will use the results provided by the authors of 
\cite{Giudice:2003jh}, i.e.\ a numerical fit to a large set of numerical 
results for $\eta$ in the MSSM for different values of $\widetilde m_1$ and 
$M_\mathrm{R1}$.

In the type I see-saw mechanism, there is a bound on the decay asymmetry, which
amounts to \cite{Hamaguchi01,Davidson:2002qv} 
\begin{eqnarray}\label{eq:Eps1BoundTypeI}
|\varepsilon_{1}| \le \frac{3 M_\mathrm{R1}}{8 \pi v_\mathrm{u}^2} 
(m_3^\mathrm{I} - m_1^\mathrm{I})  \le \frac{3 M_\mathrm{R1}}{8 \pi v_\mathrm{u}^2}\sqrt{|\Delta m^2_{\mathrm{atm}}|}
\end{eqnarray}
in the MSSM. 
This leads to a lower bound on the mass of the lightest right-handed neutrino
\cite{Davidson:2002qv}.
Assuming best efficiency, i.e.\ $\widetilde m_1$ around $10^{-3}$ eV for zero
initial population of $\nu_{\mathrm{R}1}$, the bound
is about $M_\mathrm{R1} \ge 10^9$ GeV. As we will see below, the realistic
bound in the considered classes of unified models is much higher.   
Furthermore, the bound on $M_\mathrm{R1}$ increases for increasing absolute
neutrino mass scale. This is because thermal type I leptogenesis is less 
efficient for a larger neutrino mass scale since \cite{Buchmuller:2004nz}
\begin{eqnarray}\label{eq:m1tilde_Def}
\widetilde m_1 \ge m^\mathrm{I}_{i,\mathrm{min}} \;,
\end{eqnarray}
 with $m^\mathrm{I}_{i,\mathrm{min}} :=
\mbox{min}\,(m^\mathrm{I}_1,m^\mathrm{I}_2,m^\mathrm{I}_3)$. 
Together with an improved bound 
on the type I decay asymmetry, 
this finally leads to an upper bound for the absolute mass scale of the 
light neutrinos of about $0.1$ eV \cite{Buchmuller:2003gz}.  
Let us note at this point that if neutrinoless double beta decay or a signal for
neutrino mass from direct searches is 
observed in the near future and would point to a mass above $0.1$ eV, the
requirement of successful thermal leptogenesis would disfavour the type I see-saw
mechanism, strongly pointing towards a type II see-saw.

In the case that the lightest right-handed neutrino dominates the see-saw
mechanism, $\varepsilon_{1}$ is typically proportional to   
$m_2^\mathrm{I} = \sqrt{\Delta m^2_{\mathrm{sol}}}$, a factor of
$\sqrt{m^2_{\mathrm{sol}}/|m^2_{\mathrm{atm}}|}$ smaller than the 
upper bound in Eq.~(\ref{eq:Eps1BoundTypeI}) \cite{Hirsch:2001dg}. 
In our analysis, we will however use the general bound of 
Eq.~(\ref{eq:Eps1BoundTypeI}).  
With respect to the parameter $\widetilde m_1$ which governs washout of the
produced asymmetry, from Eqs.~(\ref{eq:YnuLSD}) and (\ref{eq:m1tilde}) we now 
obtain \cite{Hirsch:2001dg}
\begin{eqnarray}\label{eq:m1tilde_LSD}
\widetilde m_1 = m_3^{\mathrm{I}}\;.
\end{eqnarray}
This implies large 
washout effects compared to its optimal value for $\widetilde m_1
\approx 10^{-3}$ eV and the efficiency 
factor $\eta$ is significantly reduced. Using the results for $\eta$ 
from \cite{Giudice:2003jh},\footnote{Note that
there are minor differences between the results quoted in the literature.
However due to the large uncertainties we allow for our estimates 
this differences are not significant for our analysis.} 
the bound on $M_\mathrm{R1}$ for a dominant lightest right-handed neutrino 
can be calculated from (combining Eqs.~(\ref{eq:nBformEpsEta}),
(\ref{eq:Eps1BoundTypeI}) and (\ref{eq:m1tilde_LSD})) 
\begin{eqnarray}\label{eq:MR1boundLSD}
M_\mathrm{R1} \ge \frac{8 \pi v_\mathrm{u}^2}{3} \,\frac{n_\mathrm{B} /n_\gamma }{1.04 \cdot 10^{-2}\, 
\sqrt{|\Delta m^2_{\mathrm{atm}}|}} \,\frac{1}{\eta} \gtrsim 10^{11}\: \mbox{GeV}\; , 
\end{eqnarray} 
where we have used the present best-fit values 
$\Delta m^2_{\mathrm{atm}} = 2.2 \cdot 10^{-3}$ eV$^2$ \cite{Maltoni:2004ei} and
 $n_\mathrm{B} /n_\gamma 
= 6.5 \cdot 10^{-10}$ \cite{Spergel:2003cb} and 
where $\eta$ in 
Eq.~(\ref{eq:MR1boundLSD}) is calculated with
$\widetilde m_1 = m_3^\mathrm{I} = \sqrt{\Delta m^2_{\mathrm{atm}}}$, 
yielding $\eta \approx 0.003$ \cite{Giudice:2003jh}. 
A similar conclusion has been
obtained in \cite{Ibarra:2003up}, where leptogenesis  
with two right-handed neutrinos and a texture zero in
the (0,0)-entry of $Y_\nu$ has been analyzed.  

The bound of Eq.~(\ref{eq:MR1boundLSD}) is clearly in conflict 
with the range $M_\mathrm{R1} \sim 2 \cdot 10^6 \:\mbox{GeV}\; ... 
\; 1 \cdot 10^9\:\mbox{GeV}$
(see Eq.~(\ref{eq:EsimatedRangeForMR1}))  
estimated for the class of unified models discussed above. 
As briefly discussed above, proposed solutions to this potential problem within
the framework of the leptogenesis mechanism might make use of
non-thermal leptogenesis via the decay of the inflaton 
\cite{nonthermalLG}. 
On the other hand, resonant leptogenesis \cite{Flanz:1996fb} does not seem to
appear natural in the considered scenario, but might well be applied to other
classes of unified flavour models \cite{Albright:2003xb}.   
Here our preferred 
route towards resolving the above conflict is to generalize the type I
see-saw mechanism to a type II see-saw. 
In the following section, we will show how raising the absolute 
neutrino mass scale by adding a type II see-saw contribution proportional 
to the unit matrix can resolve the 
conflict between the leptogenesis bound and the
prediction for $M_\mathrm{R1}$.

\section{Leptogenesis in Unified Theories with Type II See-Saw}
\label{sec:TypeII}
Extending the type I see-saw \cite{seesaw} to a type II see-saw 
mechanism \cite{seesaw2} by an additional
direct mass term for the light left-handed neutrinos has interesting
consequences for leptogenesis 
\cite{Hambye:2003ka,Antusch:2004xy,TypeIILGModels}. 
The type II see-saw mechanism also opens up new possibilities
for constructing models of fermion masses and mixings.
We have previously shown \cite{Antusch:2004xd} how 
adding a type II contribution proportional to the unit matrix  
to the neutrino mass matrix, $m_{\mathrm{LL}} = - v_\mathrm{u}^2 Y_\nu M_\mathrm{RR}^{-1}Y_\nu^T + 
m^\mathrm{II}e^{\I \delta_\Delta} \mathbbm{1}$, 
allows models with hierarchical neutrino masses to be transformed into 
type II see-saw models with a partially degenerate mass spectrum in a natural 
way.  
Schematically, the structure of the
neutrino mass matrix is given by
\begin{eqnarray}
m^\nu_{\mathrm{LL}} \approx m^{\mathrm{II}}e^{\I \delta_\Delta} 
\left(\begin{array}{ccc}
1&0&0\\
0&1&0\\
0&0&1
\end{array}\right)
 +
 \left(\begin{array}{ccc}
(m^{\mathrm{I}}_{\mathrm{LL}})_{11} &(m^{\mathrm{I}}_{\mathrm{LL}})_{12} & (m^{\mathrm{I}}_{\mathrm{LL}})_{13}\\
(m^{\mathrm{I}}_{\mathrm{LL}})_{21} &(m^{\mathrm{I}}_{\mathrm{LL}})_{22} &(m^{\mathrm{I}}_{\mathrm{LL}})_{23}\\
(m^{\mathrm{I}}_{\mathrm{LL}})_{31} &(m^{\mathrm{I}}_{\mathrm{LL}})_{32} &(m^{\mathrm{I}}_{\mathrm{LL}})_{33}
\end{array}\right).
\end{eqnarray}  
Such a ``type II upgrade'' of hierarchical type I see-saw models
has been analyzed systematically in \cite{Antusch:2004xd} 
and classes of type II see-saw 
models have been proposed which use SO(3) flavour symmetry and a real 
vacuum alignment. 
The type II part proportional to the unit matrix governs the neutrino mass 
scale, whereas the hierarchical type I part controls the neutrino mass splittings
and the mixing angles, e.g.\ using sequential dominance
\cite{SequentialD} within the type I see-saw contribution.
We will now discuss how lifting the absolute 
neutrino mass scale by adding a type II see-saw contribution 
$m^\mathrm{II} \mathbbm{1}$
 can solve the potential conflict between the prediction for $M_\mathrm{R1}$ 
 and the lower bound from leptogenesis.  
As in the previous section, we will focus on 
classes of unified flavour models where the
lightest right-handed neutrino dominates the type I see-saw contribution as an
example. In the next sub-section we shall show that the
lightest right-handed neutrino mass increases with increasing
type II neutrino mass scale. Then in the following
sub-section we shall show that the thermal leptogenesis lower bound 
on the lightest right-handed neutrino mass
decreases with increasing type II mass scale.
The combination of these two effects then resolves the 
right-handed neutrino mass conflict
between unified theories and thermal leptogenesis 
for a sufficiently high type II neutrino mass scale, which we
shall subsequently estimate.

\subsection{Right-handed Neutrino Masses and Type II See-saw}
In the following, we will use real alignment of the 
SO(3) breaking vacuum as in 
\cite{Barbieri:1999km,Antusch:2004xd,ChargedLMixing} 
for definiteness, 
which results in a specific phase
structure for the Yukawa matrices. The neutrino Yukawa matrix then has the form  
\begin{eqnarray}\label{eq:YnuLSDII}
Y_\nu =
\begin{pmatrix}
0         & a\eps^3 e^{\I \delta_2}& * \\
e\eps^3 e^{\I \delta_1}   & b\eps^3  e^{\I \delta_2}& * \\ 
f\eps^3 e^{\I \delta_1}  & c\eps^3  e^{\I \delta_2}&   {\cal O}(1)\\
\end{pmatrix},
\end{eqnarray} 
similar to Eq.~(\ref{eq:YnuLSD}), however now $e,f,q,b,c$ 
are real (not necessarily positive)  
parameters and $\delta_1,\delta_2$ are common phases for each column of
$Y_\nu$. 

For example a Pati-Salam unified model 
based on $SO(3)$ with a neutrino Yukawa matrix
similar to Eq.~(\ref{eq:YnuLSDII}) 
has recently been proposed \cite{King:2005bj}.
In the proposed model a vacuum alignment
with $a=b=c$ and $e=-f$ is used to give tri-bimaximal
neutrino mixing, but results in zero type I
leptogenesis\cite{King:2005bj}. Such a model
may in principle be ``up-graded'' to a type II model along the
lines discussed here, allowing successful leptogenesis.
This provides a good example of the type of model to which 
the results presented here may be applied.
However, models based on SU(3) \cite{King:2001uz}
cannot similarly be ``up-graded''.

Note that in realistic models the phase structure 
in Eq.~(\ref{eq:YnuLSDII}) may be 
modified by correction from higher-dimensional, next-to-leading operators. The
entries marked with a star are much smaller than 1 and do not play any role in
our analysis. 
We will
furthermore make use of the fact that in the considered class of models, only
small corrections to the neutrino mixings, compared to the present
experimental uncertainties, arise from the charged lepton sector.  
We will neglect these corrections in the following since
they only contribute marginally to the uncertainties for the estimates of 
$M_\mathrm{R1}$ and do not 
effect the leptogenesis bounds.   

Using the sequential dominance conditions 
in Eq.~(\ref{Eq:SequentialD}) 
for the type I contribution to the
neutrino mass matrix and approximating $m_1^{\mathrm{I}}=0$, 
the total masses of the light neutrinos, the eigenvalues of
$m^\nu_{LL} = m^\mathrm{I}_{LL}+m^\mathrm{II}_{LL}$, are given by    
 \begin{subequations}\label{eq:NuMassEigenvalues}\begin{eqnarray}
m_1 &\approx& m^\mathrm{II}\; , \\
m_2 &\approx& | m^{\mathrm{II}} e^{\I \delta_\Delta}- m_2^{\mathrm{I}} \,e^{i 2 \delta_2}|\; , \\
\label{eq:NuMassEigenvalues_m3} m_3 &\approx& | m^{\mathrm{II}} e^{\I \delta_\Delta}- m_3^{\mathrm{I}} 
\,e^{i 2 \delta_1}|\;, 
\end{eqnarray} \end{subequations} 
where 
$\{0,m^\mathrm{I}_2, m^\mathrm{I}_{3}\}$ are the approximate 
mass eigenvalues of the type I contribution to the neutrino
mass matrix, $m^\mathrm{I}_{\mathrm{LL}}$, and  $m^\mathrm{II}$ is defined
to be positive.

We can now calculate analytically how the mass of $M_\mathrm{R1}$ depends on
$m^\mathrm{II}$, which is equal to the mass of the lightest left-handed neutrino for a
normal mass ordering. Let us therefore first extract $m_3^\mathrm{I}$. 
Clearly, since $m_3^\mathrm{I}$ generates the mass splitting of $m_3$ and $m_1$,
for given $|\Delta m^2_\mathrm{atm}| := |m_3^2 - m_1^2|$ it has to decrease if the absolute neutrino
mass scale is lifted via $m^\mathrm{II}$. 
From Eqs.~(\ref{eq:NuMassEigenvalues_m3}), we obtain
\begin{eqnarray}\label{eq:Estimatem3I}
m_3^\mathrm{I} =
m^\mathrm{II} \cos (2\delta_1 - \delta_\Delta) \pm 
\sqrt{[m^\mathrm{II} \cos (2\delta_1 - \delta_\Delta)]^2 \pm  
|\Delta m^2_\mathrm{atm}|}\; ,
\end{eqnarray}
where the '$+$' stands for normal ordering of the mass eigenvalues, i.e.\ 
 $\cos(2\delta_1 - \delta_\Delta) < 0$, and the '$-$' stands for an inverse
 ordering corresponding to $\cos(2\delta_1 - \delta_\Delta) > 0$ (if a
solution exists which is obviously not guaranteed in the latter case for small
$m^{\mathrm{II}}$). A graphical illustration can be found in Fig.~3 of 
Ref.~\cite{Antusch:2004xd}. 
Assuming a normal mass ordering, for $[ m^{\mathrm{II}}
 \cos(2\delta_1 - \delta_\Delta)]^2 \gg |\Delta m^2_\mathrm{atm}|$, 
we obtain 
\begin{eqnarray}\label{eq:Estimatem3IforLargemII}
m_3^\mathrm{I} \sim \frac{\Delta m^2_\mathrm{atm}}{- 2 m^{\mathrm{II}} \cos
(2\delta_1 - \delta_\Delta)}\; ,
\end{eqnarray}
which shows that the type I mass contributions
(which govern the neutrino mass splittings in this approach)
decrease with increasing type II neutrino mass scale. Finally, 
from $m_3^\mathrm{I}$, 
$M_\mathrm{R1}$ is given by  
\begin{eqnarray}
M_\mathrm{R1} = \frac{(e \eps^3)^2 v_\mathrm{u}^2}{s_{23}^2 m^\mathrm{I}_3} 
\; ,
\end{eqnarray}
analogous to Eq.~(\ref{eq:MR1_LSD}).  However, compared to the type I
case, $m^\mathrm{I}_3$ can now be significantly smaller than
$\sqrt{|\Delta m^2_\mathrm{atm}|}$ for $m^{\mathrm{II}}$ close to the
present bounds for the absolute neutrino mass scale. Thus
the prediction for $M_\mathrm{R1}$ increases with increasing
type II neutrino mass scale, as claimed earlier.
We estimate that, for
$m^{\mathrm{II}}=0.2$ eV, $m_3^\mathrm{I}$ is reduced from about
$0.05$ eV to $0.006$ eV and thus the prediction for $M_\mathrm{R1}$
increases by about an order of magnitude.

\subsection{Leptogenesis Bound on $M_\mathrm{R1}$ and Type II See-Saw}
In the class of models under consideration, if the lightest
right-handed neutrino dominates the type I see-saw contribution to the
neutrino mass matrix, thermal leptogenesis becomes more efficient when
the type II neutrino mass scale increases in two ways: due to an enhanced
decay asymmetry and due to reduced washout leading to a larger
efficiency factor $\eta$.  Let us now discuss these points in
detail. They both result in a decrease of the lower bound on
$M_\mathrm{R1}$ from thermal leptogenesis.

The decay asymmetry in the type II see-saw, which generalizes the type I
decay asymmetry of Eq.~(\ref{eq:eps1typeI}), is given by \cite{Hambye:2003ka,Antusch:2004xy}
\begin{eqnarray}\label{eq:eps1typeI}
\varepsilon_{1}= 
 \frac{3}{8 \pi} 
 \frac{M_{\mathrm{R}1}}{v_\mathrm{u}^2} 
 \frac{\sum_{fg}\mbox{Im}\, [(Y^*_\nu)_{f1} (Y^*_\nu)_{g1} 
 (m_{\mathrm{LL}}^\mathrm{I} + m_{\mathrm{LL}}^\mathrm{II})_{fg}]}{(
 Y_\nu^\dagger Y_\nu)_{11}}
\end{eqnarray} 
in the limit that the lightest right-handed neutrino is much lighter than the
additional particles associated with the type I and 
type II see-saw mechanism (e.g.\
much lighter than the SU(2)$_\mathrm{L}$-triplet Higgs fields).  
It is bounded from above by \cite{Antusch:2004xy}
\begin{eqnarray}\label{eq:TypeIIBoundOnEps1}
|\varepsilon_{1}| \le \frac{3 M_\mathrm{R1}}{8 \pi v_\mathrm{u}^2}
m^\nu_{i,\mathrm{max}}\;,
\end{eqnarray}
 with $m^\nu_{i,\mathrm{max}} :=
\mbox{max}\,(m_1,m_2,m_3)$. Type I and type II bounds on $\varepsilon_{1}$ are
identical for a hierarchical neutrino mass spectrum \cite{Hambye:2003ka,Antusch:2004xy}. 
However, if the neutrino mass scale 
$m^\mathrm{II}$ increases, the type II bound increases \cite{Antusch:2004xy} 
whereas the type I bound decreases \cite{Davidson:2002qv}.

\begin{figure}[thb]
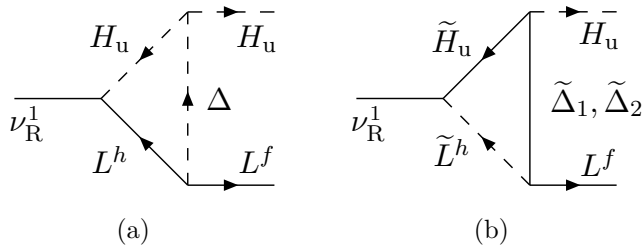

\centering
\CenterEps[1.0]{TripletDiagrams}
 \caption{\label{fig:TripletDiagrams}
 Loop diagrams in the MSSM involving virtual SU(2)$_{\mathrm{L}}$-triplets, 
 which contribute to the decay 
 $\nu^1_{\mathrm{R}}\rightarrow L^f_a H_\mathrm{u}{}_b$ in the  
 type II see-saw mechanism. 
 $\Delta$ in Fig.~\ref{fig:TripletDiagrams}(a) is the 
 SU(2)$_{\mathrm{L}}$-triplet Higgs coupling to the lepton doublets and 
 $\widetilde \Delta_1$ and $\widetilde \Delta_2$ in
 Fig.~\ref{fig:TripletDiagrams}(b) are the mass eigenstates
 corresponding to the superpartners of the 
 SU(2)$_{\mathrm{L}}$-triplet scalar fields $\Delta$ and $\bar{\Delta}$ (see
 e.g.\ \cite{Antusch:2004xy} for details).
 }
\end{figure}

Explicitly, if we add a type II see-saw contribution proportional to the unit
matrix to the class of type I see-saw models under consideration, we 
 obtain 
\begin{eqnarray}\label{eq:epa1inTypeIILSD}
\varepsilon_{1} = - \frac{3 M_\mathrm{R1}}{8 \pi v_\mathrm{u}^2} 
\left[ \sin (2 \delta_1 - \delta_\Delta) m^\mathrm{II} \pm {\cal
O}(m_2^\mathrm{I}) \right].
\end{eqnarray}
When increasing the absolute neutrino mass scale, $m_2^\mathrm{I}$ decreases very fast
(see e.g.\ Fig.~5(a) of Ref.~\cite{Antusch:2004xd}) and the decay asymmetry is 
typically dominated by the type II contribution 
already for $m^\mathrm{II}$ larger than about $0.03$ eV. We note
that the bound on $\varepsilon_{1}$ 
of Eq.~(\ref{eq:TypeIIBoundOnEps1}) can be nearly saturated 
with a type II see-saw contribution proportional to the unit
matrix in a natural way. 
If the type II contribution to the decay asymmetry 
dominates leptogenesis, the ``leptogenesis phase'' in our scenario is given by
\begin{eqnarray}
\delta_{\mathrm{cosm}} = 2 \delta_1 - \delta_\Delta \;.
\end{eqnarray}
The decay asymmetry dominantly stems from the interference of the 
tree-level decay of $\nu_{\mathrm{R1}}$ with the one-loop diagram where 
the triplet responsible for the type II see-saw contribution 
or its superpartner run in the loop  
 (see Fig.~\ref{fig:TripletDiagrams}). 
It is interesting to note that although classes of ``type-II-upgraded'' 
see-saw models studied in \cite{Antusch:2004xd} 
have the generic property that all low energy observable CP phases 
from the neutrino sector become smaller as the neutrino mass scale 
increases 
(e.g.\ the Dirac CP phase $\delta$ observable in neutrino oscillations), the
phase $\delta_{\mathrm{cosm}}$ relevant for leptogenesis is unaffected
and remains finite in the large type II mass limit.

In type II leptogenesis with $M_\mathrm{R1}$ much lighter than
other contributions to the see-saw mechanism, the efficiency factor $\eta$ 
is typically still determined by $M_\mathrm{R1}$ and the Yukawa couplings to 
$\nu_{\mathrm{R1}}$ \cite{Hambye:2003ka} and, in particular, washout effects  
from $\Delta L \!=\! 2$-scattering processes involving 
the SU(2)$_\mathrm{L}$-triplets are negligible for $M_\Delta 
\gg M_\mathrm{R1}$.\footnote{One might argue that the contribution to
washout from $\Delta L \!=\! 2$-scattering processes involving the heavy 
SU(2)$_\mathrm{L}$-triplet $\Delta$ can be treated analagously to the
contribution from the heavy right-handed
neutrinos $\nu_{\mathrm{R}2},\nu_{\mathrm{R}3}$. For $M_\Delta ,
M_\mathrm{R2},M_\mathrm{R3} \gg M_\mathrm{R1}$ these heavy fields can be
effectively integrated out, contributing via the same effective dimension 5 neutrino 
mass operator. The $\Delta L \!=\! 2$-scattering processes can then be
neglected for $M_\mathrm{R1} \ll 10^{14}\,\mbox{GeV}\:
(0.05\,\mbox{eV}\:/\overline{m})^2$ \cite{DiBari:2004en}, with $\overline{m}^2 := m_1^2+m_2^2+m_3^2$,
where this is also valid in the type II see-saw case. 
}   
In the following, we will assume that 
to a good approximation $\eta$ still 
depends only on $\widetilde m_1$ \cite{Buchmuller:2000as}, 
defined in Eq.~(\ref{eq:m1tilde_Def}),  
in the same way as in
the type I see-saw mechanism.
For our estimates, we will use the results for $\eta(\widetilde m_1)$ 
of \cite{Giudice:2003jh}. 
In the scenario under consideration, 
\begin{eqnarray}
\widetilde m_1 = m_3^\mathrm{I} \; , 
\end{eqnarray}
which means it decreases if the neutrino mass scale
is lifted via $m^\mathrm{II}_\mathrm{LL} = m^\mathrm{II} \mathbbm{1}$ (cf.\ 
Eqs.~(\ref{eq:Estimatem3I}) and (\ref{eq:Estimatem3IforLargemII})). 
Quantitatively, if we assume a
neutrino mass scale $m^\mathrm{II}=0.2$ eV, we see from 
Eq.~(\ref{eq:Estimatem3IforLargemII}) that $m_3^\mathrm{I}$ reduces from about 
$0.05$ eV to $0.006$ eV, leading to an increase of $\eta$
from about $0.003$ to $0.04$. Note that for $\widetilde m_1 =0.006$ eV, 
$\eta$ is still nearly independent of the initial population of right-handed
neutrinos (see e.g.\ \cite{Giudice:2003jh}). 

Using Eqs.~(\ref{eq:nBformEpsEta}), (\ref{eq:m1tilde_LSD}), 
(\ref{eq:Estimatem3I}) and 
(\ref{eq:epa1inTypeIILSD}), 
the lower bound on the mass of the right-handed neutrino from the
requirement of successful thermal leptogenesis can be calculated from  
 \begin{eqnarray}\label{eq:MR1boundLSDTypeIIp}
M_\mathrm{R1} \ge \frac{8 \pi v_\mathrm{u}^2}{3} \,
\frac{n_\mathrm{B} /n_\gamma }{1.04 \cdot 10^{-2}\, 
[\sin (2 \delta_1 - \delta_\Delta) \,m^\mathrm{II} + m_3^\mathrm{I}]} \,
\frac{1}{\eta} \; , 
\end{eqnarray} 
which now depends on $m^\mathrm{II}$. Note that $\eta$ in 
Eq.~(\ref{eq:MR1boundLSDTypeIIp}) is calculated with
$\widetilde m_1 = m_3^\mathrm{I}$. The lower bound on $M_\mathrm{R1}$   
thus decreases with increasing neutrino mass due to the explicit factor $m^\mathrm{II}$ in the
denominator (from the decay asymmetry) and due to an increase in $\eta$.  

\subsection{Numerical Results}

\begin{figure}[p]
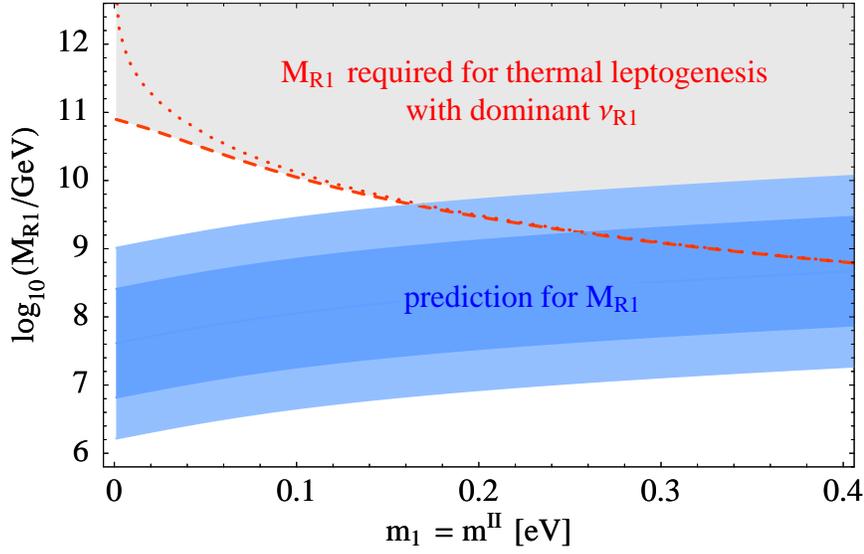
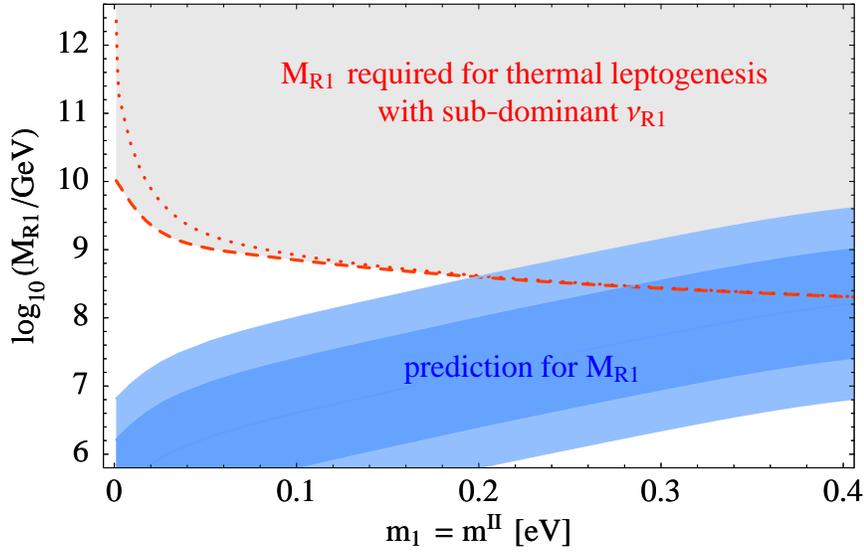

\begin{center}
  \subfigure[]{$\CenterEps[1]{LG_TypeII_LSD}$}\\
  \subfigure[]{$\CenterEps[1]{LG_TypeII_ISD}$} 
 \caption{\label{fig:LG_TypeII}
 Estimates for the mass of the lightest right-handed neutrino $M_\mathrm{R1}$ 
in unified theories with type II see-saw, compared to the lower bounds from 
successful thermal leptogenesis
(dashed line) with $M_\mathrm{R1} \ll M_\mathrm{R2},M_\mathrm{R3},
M_\mathrm{\Delta}$. Fig.~\ref{fig:LG_TypeII}(a) shows the results for classes of
unified models where the lightest right-handed neutrino dominates the type I
see-saw contribution and Fig.~\ref{fig:LG_TypeII}(b) shows the results where it
is sub-dominant. 
The dotted lines are the leptogenesis 
bounds on $M_\mathrm{R1}$ with 
$\varepsilon_1^\mathrm{I}$ set to zero. 
Note that in the type I limit where the neutrino mass scale 
$m^\mathrm{II}=0$ is zero, the leptogenesis bounds 
are more stringent than the general bound 
 $\sim 10^9$ GeV due to larger washout and clearly in conflict with the
 predictions for $M_\mathrm{R1}$. 
 However, the bounds decrease with
 increasing neutrino mass scale and in addition the predictions for $M_\mathrm{R1}$
 increase, allowing for consistent thermal leptogenesis 
if neutrino masses are larger than about $0.15$ eV. 
 }
\end{center}
\end{figure}

We have seen that adding a type II contribution proportional to the
unit matrix, leads to an increase in the
prediction for $M_\mathrm{R1}$ in the considered class of unified flavour models and in addition 
to a decrease of the lower bound on $M_\mathrm{R1}$ from the requirement of
successful thermal leptogenesis. Quantitatively, this is shown in
Fig.~\ref{fig:LG_TypeII}(a) for a leptogenesis phase chosen to be 
$\delta_{\mathrm{cosm}} = 2 \delta_1 - \delta_\Delta=135^\circ$. In addition, we
have set the type I contribution $\epsilon_1^\mathrm{I}$ to the decay asymmetry
to its maximal value proportional to $m_3^\mathrm{I}$, such that in the $m^\mathrm{II}=0$ 
limit we obtain the type I bound. We have also used the same range for
$e \eps^3$ as in the discussion of the type I see-saw models. 
RG effects \cite{betakappa}
are included for  
$\tan \beta = 10$ as an example, using the software packages REAP/MPT 
introduced in \cite{Antusch:2005gp}.
We find that in unified theories where the lightest right-handed neutrino
dominates the see-saw mechanism, thermal leptogenesis is possible if  
$m^\mathrm{II}$ is larger than about $0.15$ eV.

\section{Leptogenesis in Unified Theories with a \\ Sub-Dominant
Lightest RH Neutrino}\label{App:A}

In this section we relax the assumption that the lightest
right-handed neutrino dominates the see-saw mechanism,
and briefly discuss leptogenesis and right-handed neutrino masses in unified
theories where the lightest right-handed neutrino is sub-dominant within 
the type I see-saw contribution. 
To be precise we shall assume in this section that the lightest
right-handed neutrino is mainly responsible for the 
type I contribution to the solar neutrino mass, while the 
next-to-lightest right-handed neutrino is mainly responsible for the
type I contribution to the atmospheric neutrino mass.
This is sometimes referred to as intermediate sequential dominance
(ISD) \cite{SequentialD}. For the neutrino Yukawa matrix, we assume the form  
\begin{eqnarray}\label{eq:YnuISD}
Y_\nu \sim
\begin{pmatrix}
a\eps^4   e^{\I \delta_1} & *       & * \\
b\eps^4   e^{\I \delta_1} & e\eps^2  e^{\I \delta_2}& * \\
c\eps^4    e^{\I \delta_1}& f\eps^2  e^{\I \delta_2}& {\cal O}(1)  \\
\end{pmatrix} ,
\end{eqnarray} 
where $\eps_\nu = \eps_\mathrm{u} \approx 0.05$  
and where the entries marked
with a star are much smaller than the other entries in the corresponding
column of $Y_\nu$, where the RH neutrino associated with the 
second column dominates the see-saw
mechanism, and the first column gives the leading sub-dominant contributions.
An important feature is that $Y_\nu$ is
linked to the quark Yukawa matrix $Y_\mathrm{u}$, so that 
$(Y_\nu)_{11}=a\eps^4$ 
is related to the up-quark Yukawa coupling which can be estimated as  
$y_\mathrm{u} \approx (Y_\mathrm{u})_{11} \approx 4.7 \cdot 10^{-6}$ at the GUT scale
(see e.g.\ \cite{King:1998nv}). We will use the range $a\eps^4 \in [\tfrac{1}{5},5] \cdot  
4.7 \cdot 10^{-6}$ in our analysis. 
The mass of the lightest right-handed neutrino is then given by 
\begin{eqnarray}\label{eq:MR1_ISD}
M_\mathrm{R1} \;=\; \frac{(a \eps^4)^2 v_\mathrm{u}^2}{(s^\nu_{12})^2 
m^\mathrm{I}_2} 
\quad \mbox{where} \quad  
m_2^\mathrm{I} \;\sim\; \frac{\Delta m^2_\mathrm{sol}}{- 2 m^{\mathrm{II}} \cos
(2\delta_1 - \delta_\Delta)} \; ,
\end{eqnarray}
in the limit of large $m^\mathrm{II}$, analogous to 
Eqs.~(\ref{eq:Estimatem3I}) and (\ref{eq:Estimatem3IforLargemII}). 
$\Delta m^2_\mathrm{sol}$ is defined in the usual way as $m_2^2 - m_1^2$.  
We see that $m_2^\mathrm{I}$ decreases
with increasing neutrino mass scale, even faster than $m_3^\mathrm{I}$, and thus
$M_\mathrm{R1}$ increases significantly. In the type I limit where $m_2^\mathrm{I}=\sqrt{\Delta m^2_\mathrm{sol}}$, 
$M_\mathrm{R1}$ is predicted to be in the range 
\begin{eqnarray}
M_\mathrm{R1} \;\sim \; 2 \cdot 10^4 \:\mbox{GeV} \; ... \; 
7 \cdot 10^6\:\mbox{GeV}\;,
\end{eqnarray}
clearly incompatible with requirements on $M_\mathrm{R1}$ from thermal leptogenesis. 
For $\widetilde m_1$ we find 
\begin{eqnarray}
\widetilde m_1 \ge m_2^\mathrm{I}
\end{eqnarray} 
and the lower bound on the mass of the right-handed neutrino from the
requirement of successful thermal leptogenesis can be calculated from  
 \begin{eqnarray}\label{eq:MR1boundLSDTypeII}
M_\mathrm{R1} \ge \frac{8 \pi v_\mathrm{u}^2}{3} \,
\frac{n_\mathrm{B} /n_\gamma }{1.04 \cdot 10^{-2}\, 
[\sin (2 \delta_1 - \delta_\Delta) \,m^\mathrm{II} + m_3^\mathrm{I})]} \,
\frac{1}{\eta} \; . 
\end{eqnarray} 
Note that in Eq.~(\ref{eq:MR1boundLSDTypeII}), $\eta$ is calculated with 
$\widetilde m_1 = m_2^\mathrm{I}$. 
In the type I limit, we obtain a lower bound for $M_\mathrm{R1}$ of about 
$10^{10}$ GeV. 
This bound decreases with increasing neutrino mass scale $m^\mathrm{II}$ since
washout can be smaller for lower $m_2^\mathrm{I}$ and since the decay asymmetry
increases with $m^\mathrm{II}$. 

 The prediction for the mass of the lightest right-handed neutrino is compared to
 the numerical results for the lower bound from successful thermal leptogenesis in
 Fig.~\ref{fig:LG_TypeII}(b). We find that for the example with $\tan \beta = 10$, a
 neutrino mass scale larger than about $0.2$ eV is required for consistent 
 thermal leptogenesis, assuming zero initial 
 population of right-handed neutrinos. 
 We note that for sub-dominant $\nu_\mathrm{R1}$ and $m^\mathrm{II}$ larger
 than about $0.03$ eV, $\widetilde m_1$ can be smaller than $\approx 10^{-3}$ and 
 the efficiency factor $\eta$ can depend on initial conditions. 
 On the contrary,
 with the lightest right-handed neutrino being dominant in the see-saw mechanism 
 we have found that for masses up to about $0.2$ eV, thermal leptogenesis is
 still in the strong washout regime and the produced baryon asymmetry is 
 nearly independent of initial conditions.
 Furthermore, with quasi-degenerate neutrino masses RG running of the neutrino 
 parameters 
 between low energy and $M_\mathrm{R1}$ has to be taken into account carefully, 
 in particular for the mixing angle 
 $\theta_{12}$ entering Eq.~(\ref{eq:MR1_ISD}) and for the solar mass squared 
 difference.\footnote{For 
 quasi-degenerate neutrino masses, the running of $\theta_{12}$ is generically 
 much stronger than the running of the other mixing angles 
 \cite{Antusch:2002hy}, 
 in particular with a dominant type II contribution proportional to the unit matrix,
 which implies small Majorana CP phases \cite{Antusch:2004xd}.}   
Due to RG effects, the required value of $m^\mathrm{II}$ may 
vary for different choices of $\tan \beta$, however this does not change the 
general result that a non-zero 
type II contribution is required.

\section{Summary and Conclusions}
As pointed out by many authors, in some classes of unified theories
where the neutrino Yukawa matrix is linked to the up-quark Yukawa
matrix the prediction for the mass of the lightest right-handed
neutrino is in conflict with the lower bound from the requirement of
successful thermal leptogenesis.  In this study, we have investigated
how lifting the absolute neutrino mass scale by adding a type II
see-saw contribution proportional to the unit matrix can resolve this
potential problem.  We found that in these classes of type II see-saw
models, lifting the neutrino mass scale increases the predictions for
the masses of the right-handed neutrinos while the decay asymmetry for
leptogenesis is enhanced and washout effects are reduced, thereby relaxing the
lower bound on the mass of the lightest right-handed neutrino from
thermal leptogenesis.  A type II see-saw contribution proportional to
the unit matrix can be realized using for instance SO(3) family
symmetry or discrete symmetries. It provides a natural way of
transforming a type I see-saw model for hierarchical neutrino masses
into a type II see-saw model for quasi-degenerate neutrinos.

We have mainly focussed on classes of unified theories where the
lightest right-handed neutrino dominates the type I see-saw mechanism,
and where sequential dominance provides a natural mechanism for giving a
neutrino mass hierarchy and bi-large neutrino mixing angles in the
presence of small charged fermion mixing angles.  We have shown
that in this type I see-saw scenario for hierarchical neutrino masses,
the prediction for the mass of the lightest right-handed neutrino is
in conflict with the lower bound from the requirement of successful
thermal leptogenesis.  We have then discussed in detail how lifting the
absolute neutrino mass scale by adding a type II see-saw contribution
proportional to the unit matrix can resolve this conflict.  We have
found that thermal leptogenesis becomes possible with a neutrino mass
scale larger than about $0.15$ eV, which implies observable 
neutrinoless double beta decay 
(and possibly also a signal from direct neutrino
mass searches) in the near future.  For such neutrino masses, thermal
leptogenesis remains in the so-called strong washout regime, where the
produced baryon asymmetry is virtually independent of initial
conditions.

We have also discussed 
classes of unified models where the second lightest right-handed neutrino
dominates the type I see-saw mechanism, and the lightest provides the leading
sub-dominant contribution. In such models 
the prediction for the mass of the lightest right-handed neutrino is
also conflict with the lower bound from thermal leptogenesis, and
again this conflict may be resolved by a type II see-saw up-grade
similar to the previous case of a dominant lightest right-handed
neutrino. However,
if the heaviest right-handed neutrino dominates the see-saw mechanism,
and the lightest right-handed neutrino is effectively decoupled,
then there is generically no conflict between leptogenesis and unified 
models, but the Yukawa matrices must involve large mixings.

\section*{Acknowledgements}
We acknowledge support from the PPARC grant PPA/G/O/2002/00468.

\providecommand{\bysame}{\leavevmode\hbox to3em{\hrulefill}\thinspace}

\end{document}